\documentclass[onecolumn,showpacs,preprintnumbers,showkeywords]{revtex4}
\usepackage{amsmath}
\usepackage{graphicx}
\usepackage{dcolumn}
\usepackage{bm}
\usepackage{subfigure}
\usepackage{times}
\begin{document}
\title{Robust creation of atomic W state in a cavity by adiabatic passage }
\author{M. Amniat-Talab$^{1}$}
\email{amniyatm@u-bourgogne.fr}
\author{M. Saadati Niari$^{1}$}
\author{S. Gu\'{e}rin$^{2}$}
 \affiliation{$^{1}$Physics Department, Faculty of Sciences, Urmia
University, P.B. 165, Urmia, Iran.\\
$^{2}$Laboratoire de Physique, UMR CNRS 5027, Universit\'{e} de
Bourgogne, B.P. 47870, F-21078 Dijon, France.}
\date{\today }
\begin{abstract}
We propose two robust schemes to generate controllable
(deterministic) atomic W-states of three three-level atoms
interacting with an optical cavity and a laser beam. Losses  due
to atomic spontaneous emissions and to cavity decay are
efficiently suppressed by employing  adiabatic passage technique
and appropriately designed atom-field couplings. In these schemes
the three atoms traverse the cavity-mode and the laser beam  and
become entangled in the free space outside the cavity.
\end{abstract}
 \maketitle
\section{Introduction\label{intro}}
Quantum-state engineering , i.e., active control over the coherent
dynamics of suitable quantum-mechanical systems to achieve a
preselected state (e.g. entangled states or multi-photon field
states) of the system, has become a fascinating prospect of modern
physics. The physics of entanglement provides the basis of
applications such as quantum information processing and quantum
communications. Particles can then be viewed as carriers of
quantum bits of information and the realization of engineered
entanglement is an essential ingredient of the implementation of
quantum gates \cite{qgate}, cryptography \cite{EkertPRL91} and
teleportation \cite{BennettPRL93}. The creation of long-lived
entangled pairs of atoms may provide reliable quantum information
storage. The idea is to apply a set of controlled coherent
interactions to the atoms of the system in order to bring them
into a tailored entangled state. The problem of controlling
entanglement is thus directly connected to the problem of coherent
control of population transfer in multilevel systems.

In the case of tripartite entanglement, there are two classes of
tripartite entangled states, the GHZ class \cite{GHZ} and the W
class, which are inequivalent under stochastic local operation and
classical communication \cite{DUR}. One of the interesting
properties of W states [ such as
$\frac{1}{\sqrt{3}}(|0,0,1\rangle+|0,1,0\rangle+|1,0,0\rangle)$]
is that if one particle is traced out, there remains entanglement
of the remaining two particles, or if one particle is measured in
basis $\{|0\rangle, |1\rangle\}$, then the state of remained two
particles is either in a maximally entangled state or in a product
state. In the context of cavity QED, the stimulated Raman
adiabatic passage (STIRAP) technique  has been introduced by
Parkins et. al. \cite{ParkinsPRL93} where the Stokes pulse is
replaced by a mode of a high-Q cavity. The advantage of STIRAP is
the robustness of its control with respect to the precise tuning
of pulse areas, pulse delay, pulse widths, pulse shapes, and
detunings. In $\Lambda$-type systems, fractional STIRAP (F-STIRAP)
is a variation of STIRAP \cite{stirap} which allows the creation
of any preselected coherent superposition of the two degenerate
ground states \cite{VitanovJPB99}. The half-STIRAP process
(F-STIRAP with final half population of two ground states) has
also been studied in an optical cavity to prepare atom-photon and
atom-atom entanglement \cite{AMNPRA51,AMNPRA52}.

In cavity QED, the schemes for generating n-partite W states are
two types: 1) n-atom, and  2) n-cavity W states. The long-lived
atomic W states is more robust than the cavity W states against
the cavity damping. In this context, there are several schemes
discussing the preparation of atomic W states \cite{DENGPRA06,
GUOPRA02, CHENPRA07, BISWAS04, MarrPRA03}. The entangled W state
of two-level atoms using pulse area technique in a cavity has been
proposed in \cite{DENGPRA06, GUOPRA02}. However the pulse area
technique is not robust with respect to the velocity of the atoms
and the exact-resonance condition. Another scheme to entangle
traveling atoms in  the atom-cavity-laser system via adiabatic
passage has been proposed in Ref. \cite{MarrPRA03}. However this
scheme requires to turn off the laser field when the two atoms
have equal coupling with the cavity-mode, which is very difficult
from the experimental  point of view. Moreover this scheme
requires to compensate a dynamical Stark shift which is also very
difficult in a real experiment. In the scheme of Ref.
\cite{CHENPRA07}, $n$-atom W state is generated in $n+1$ distant
optical cavity connected by $n$ optical fibers, based on the
method proposed in \cite{AMNPRA52} and STIRAP process. The STIRAP
technique is also used in \cite{BISWAS04} to generate W state in a
two-mode optical cavity which needs initially one photon to be
present in one of the cavity modes.

In this paper we propose an alternative method to create the W
state of three traveling atoms interacting with an optical cavity
and a laser beam, based on 3-level interactions in a
$\Lambda$-configuration. This method is based on the coherent
creation of superposition of atom-atom-atom-cavity states via
fractional STIRAP and multilevel STIRAP \cite{AMNPRA52,unanyan},
that keeps the cavity-mode and the excited atomic states weakly
populated during the whole interaction in the adiabatic limit and
for a sufficiently strong cavity coupling.

\section{Construction of the model}\label{model}

Figure \ref{xy1} represents the linkage pattern of the
atom-cavity-laser system. The laser pulse associated to the Rabi frequency $%
\Omega (t)$ couples the states $|g_{1}\rangle $ and $|e\rangle$,
and the
cavity-mode  with Rabi frequency $G(t)$ couples the states $%
|e\rangle$ and $|g_{2}\rangle$. The Rabi frequencies $\Omega(t)$
  and $G(t)$ are chosen real without loss
of generality. These two fields interact with the atom with a time
delay, each of the fields is in one-photon resonance with the
respective transition. The semiclassical Hamiltonian of this
system in the resonant approximation where
\begin{equation}\label{RWA}
|\Omega_{0}|, |G_{0}|\ll \omega_{e},\omega_{C},
\end{equation}
with $\Omega_{0},G_{0}$ the peak values of the Rabi frequencies,
can then be written  as ($\hbar=1$):
\begin{eqnarray}\label{H}
    H(t)&=&\sum_{i=1,2,3}\big[\omega_{e}|e\rangle_{ii}\langle
    e|+\big(G_{i}(t)~a|e\rangle_{ii}\langle
    g_{2}|+\text{H.c.}\big)\nonumber\\
    &+&\big(\Omega_{i}(t)|g_{1}\rangle_{ii}\langle
    e|+\text{H.c.}\big)\big]+\omega _{C}a^{\dag}a,
\end{eqnarray}
where the subscript $i$ on the states denotes the three atoms, $a$
is the annihilation operator of the cavity mode, $\omega _{e}$ is
the energy of the atomic excited state $(\omega _{g_{1}}=\omega
_{g_{2}}=0)$, and $ \omega _{C}$ is the frequency of the cavity
mode taking resonant $\omega _{C}=\omega _{L}=\omega _{e}$. In the
following we consider the state of the atom1-atom2-atom3-cavity
system as $|A1,A2,A3,n\rangle$ where $\{A1,A2,A3=g_{1},e,g_{2}\}$,
and $\{n=0,1\}$ is the number of photons in the cavity-mode.
\begin{figure}
\includegraphics[width=7cm]{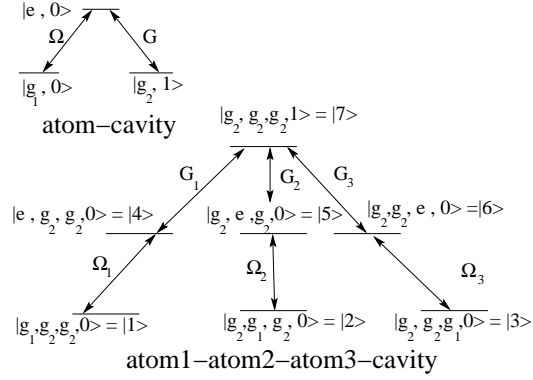}
\caption{Linkage pattern of  the  system corresponding to the
effective Hamiltonian.}\label{linkage}
\end{figure}
Regarding Figure \ref{linkage}, the subspace $\mathcal{S}$
generated by the states
$\{|g_{1},g_{2},g_{2},0\rangle$,$|e,g_{2},g_{2},0\rangle$,$|g_{2},g_{2},g_{2},1\rangle$,$|g_{2},e,g_{2},0\rangle$,$
|g_{2},g_{1},g_{2},0\rangle$,$|g_{2},g_{2},e,0\rangle$,$
|g_{2},g_{2},g_{1},0\rangle\}$ is decoupled under $H$ from the
rest of the Hilbert space of the system. If we consider the
initial state of the system as $|g_{1},g_{2},g_{2},0\rangle$ or
$|g_{2},g_{2},g_{2},1\rangle$, the effective Hamiltonian of the
system in the subspace $\mathcal{S}$ will be
\begin{equation}
    H^{\text{eff}}(t):=\left(
\begin{array}{ccccccc}
0 & 0 & 0 & \Omega_{1}(t) & 0 & 0 & 0 \\
0 & 0 & 0 & 0 & \Omega_{2}(t) & 0 & 0 \\
0 & 0 & 0 & 0 & 0 & \Omega_{3}(t) & 0 \\
\Omega_{1}(t) & 0 & 0 & 0 & 0 & 0 & G_{1}(t) \\
0& \Omega_{2}(t) & 0  & 0 & 0 & 0 & G_{2}(t) \\
0 & 0 & \Omega_{3}(t) & 0 & 0 & 0 & G_{3}(t) \\
0 & 0 & 0 & G_{1}(t) & G_{2}(t) & G_{3}(t) & 0
\end{array}%
\right),
\end{equation}
 The associated dynamics is determined by the Schr\"{o}dinger equation $i\frac{%
\partial }{\partial t}|\Phi (t)\rangle =H^{\text{eff}}(t)|\Phi (t)\rangle$.

\section{Three-atom W state}\label{Wstate}
In this section, the goal is to transform the initial state of the
system,  at the end of interaction, into an atomic W  state
\begin{eqnarray}\label{Phi-fin}
 |\Phi ( t_{f} )\rangle&=& \frac{1}{\sqrt{3}}\left( |g_{1},g_{2},g_{2},0\rangle+
 |g_{2},g_{1},g_{2},0\rangle+ |g_{2},g_{2},g_{1},0\rangle\right)\nonumber\\
 &=&\frac{1}{\sqrt{3}}\left(|g_{1},g_{2},g_{2}\rangle+
 |g_{2},g_{1},g_{2}\rangle+|g_{2},g_{2},g_{1}\rangle\right)|0\rangle,
\end{eqnarray}
where  the cavity-mode state factorizes and is left in the vacuum
state. The qubits are stored in the two degenerate ground states
of the atoms. The decoherence due to atomic spontaneous emission
is produced if the  states
$\{|e,g_{2},g_{2},0\rangle,|g_{2},e,g_{2},0\rangle\,|g_{2},g_{2},e,0\rangle\}$
are populated, and the cavity decay occurs if the  state
$|g_{2},g_{2},g_{2},1\rangle$ is populated during the adiabatic
evolution of the system. Therefore we will design the Rabi
frequencies $\{\Omega_{1}(t),G_{1}(t),\Omega_{2}(t),G_{2}(t),
\Omega_{3}(t),G_{3}(t)\}$ in our scheme such that these states are
not populated during the whole dynamics. We remark that we will
use a resonant process without any adiabatic elimination.
\subsection{Scheme 1}\label{scheme1}
\begin{figure}
\includegraphics[width=6cm]{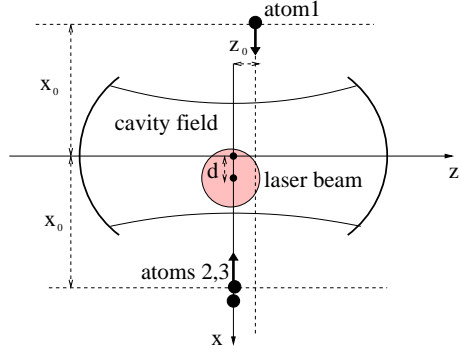}
\caption{(color online). Geometrical configuration  of the
atoms-cavity-laser system in  scheme 1. The propagation direction
of the laser beam is parallel to $y$ axis and perpendicular to the
page.}\label{xy1}
\end{figure}

In the first scheme, the system is taken to be initially in the
state $|\Phi ( t_{i} )\rangle=|g_{1},g_{2},g_{2},0\rangle$. Using
fractional STIRAP process, and Rabi frequencies of the fields as
$G_{1}(t)\sim G_{2}(t)= G_{3}(t)= G(t)$,
$\Omega_{2}(t)=\Omega_{3}(t)=\Omega(t)$, We try to transfer the
final population of the system to the desired W state. In this
case, one of the instantaneous eigenstates (the three-atom dark
state) of $H^{\text{eff}}(t) $ which corresponds to a zero
eigenvalue  is \cite{PellizzariPRL95}:
\begin{equation}\label{dark1}
|D(t)\rangle=C\Big(|g_{1},g_{2},g_{2},0\rangle
-\frac{\Omega_{1}}{G}|g_{2},g_{2},g_{2},1\rangle+
\frac{\Omega_{1}}{\Omega}|g_{2},g_{1},g_{2},0\rangle+\frac{\Omega_{1}}{\Omega}|g_{2},g_{2},g_{1},0\rangle\Big),
\end{equation}%
where $C$ is a normalization factor. The possibility of
decoherence-free generation of W state arises from the following
behavior of the dark  state:
\begin{subequations}
\begin{eqnarray}\label{beh3}
&&~t_{i}< t< t_{f},~G(t)\gg\Omega_{1} (t),\Omega(t),\\
  && |D(t)\rangle=C \big(|g_{1},g_{2},g_{2},0\rangle+\frac{\Omega_{1}}{\Omega}|g_{2},g_{1},g_{2},0\rangle+\frac{\Omega_{1}}{\Omega}
  |g_{2},g_{2},g_{1},0\rangle\big),\nonumber\\\label{beh1}
  &&\lim_{t\rightarrow t_{i}}\frac{\Omega_{1} (t)}{\Omega(t)}=0,\qquad |D(t_{i})\rangle= |g_{1},g_{2},g_{2},0\rangle\\\label{beh2}
  \nonumber\\&&\lim_{t\rightarrow t_{f} }\frac{\Omega _{1}(t)}{\Omega(t)}=1,\\
  &&
  |D(t_{f})\rangle=\frac{1}{\sqrt{3}}\big(|g_{1},g_{2},g_{2},0\rangle+|g_{2},g_{1},g_{2},0\rangle+|g_{2},g_{2},g_{1},0\rangle\big)\nonumber.
\end{eqnarray}
\end{subequations}
Equations (\ref{beh1}) and (\ref{beh2}) are known as  f-STIRAP
conditions \cite{VitanovJPB99,AMNPRA51,AMNPRA52}, and the
condition (\ref{beh3}) guarantees the absence of population in the
state $|g_{2},g_{2},g_{2},1\rangle$ during the time-evolution of
the system \cite{VitanovPRA98}. Equation (\ref{beh2}) means that
the Rabi frequencies fall off in a constant ratio, during the time
interval where they are non-negligible. We remark that this
formulation opens up the possibility to implement f-STIRAP with
Gaussian pulses. The goal in the following is to show that such a
pulse sequence can be designed in a cavity by an appropriate
choice of the parameters. In an optical cavity, the spatial
variation of the atom-field coupling for the maximum coupling
TEM$_{00}$ mode, resonant with the $|e\rangle \leftrightarrow
|g_{2}\rangle $ atomic transition, is given by
\begin{equation}
G(x,y,z)=G_{0}~e^{-(x^{2}+y^{2})/W_{C}^{2}}\cos \left( \frac{2\pi z}{\lambda }%
\right),  \label{RabiC}
\end{equation}
where $W_{L}$ is the waist of the cavity mode, and
$G_{0}=-\mu\sqrt{\omega_{C}/(2\epsilon_{0}V_{\text{mode}})}$ with
$\mu$ and $V_{\text{mode}}$  respectively the dipole moment of the
atomic transition and the effective volume of the cavity mode. The
spatial variation of the atom-laser coupling for the laser beam of
Fig. \ref{xy1} is
\begin{equation}\label{RabiL}
    \Omega(x,z)=\Omega_{0}~e^{-(x^{2}+z^{2})/W_{L}^{2}},
\end{equation}
where $W_{L}$ is the waist of the laser beam, and
$\Omega_{0}=-\mu\mathcal{E}/2$ with $\mathcal{E}$ the  amplitude
of the laser field. Figure \ref{xy1} shows a situation where the
first atom, initially in the state $|g_{1}\rangle $, goes with
velocity $v$ (on the $y=0$ plane at $z=z_{0}$ line) through an
optical cavity initially in the vacuum state $|0\rangle$ and then
encounters the laser beam, which is parallel to the $y$ axis
(orthogonal to the cavity axis and the trajectory of the atom).
The laser beam is resonant with the $|e\rangle \leftrightarrow
|g_{1}\rangle$ transition. The distance between the center of the
cavity and the laser axis is $d$. The second and the third atoms,
initially in the state $|g_{2}\rangle $ and synchronized with the
first one, move with the same velocity $v$ on the $y=0$ plane at
$z=0$ in the opposite direction with respect to the first atom.
The traveling atoms encounter the time-dependent and delayed Rabi
frequencies of the cavity-mode and the laser fields as follows:
\begin{subequations}
\begin{eqnarray}
G_{1}(t) &=&G_{0}~e^{-(vt)^{2}/W_{C}^{2}}\cos \left( \frac{2\pi z_{0}}{\lambda }%
\right) , \\
\Omega_{1}(t)
&=&\Omega_{0}~e^{-z_{0}^{2}/W_{L}^{2}}~e^{-(vt-d)^{2}/W_{L}^{2}},\\
G_{2}(t) &=& G_{3}(t)=G_{0}~e^{-(vt)^{2}/W_{C}^{2}},\\
\Omega_{2}(t)&=&\Omega_{3}(t)=\Omega_{0}~e^{-(vt+d)^{2}/W_{L}^{2}}
\end{eqnarray}
\end{subequations}

where the time origin is defined when the atoms meet the center of
the cavity at $x=y=0$.
\begin{figure}
\includegraphics[width=7cm]{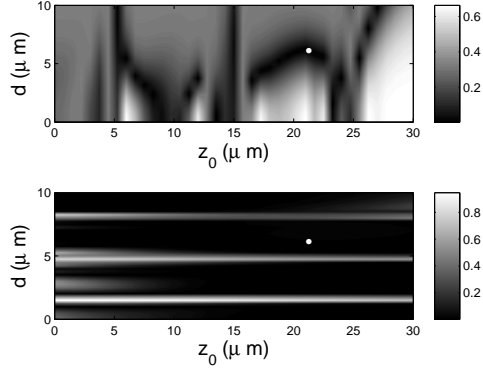}
\caption{Top panel: contour plot at the final time $t_{f}$ of $|\frac{1}{3}%
-P_{|g_{1},g_{2},g_{2},0\rangle }(t_{f})|$ as a function of
$z_{0}$ and $d$ (black areas correspond to approximately $33\%$
population transfer) with the pulse
parameters as $W_{L}=20~\mu$m, $W_{C}=40~\mu$m, $v=2$ m/s,$~%
\lambda =780$ nm, $~\Omega _{0}=20(v/W_{L}),~G_{0}=100(v/W_{C})$.
Bottom panel: The same plot for the sum of the final populations
in intermediate states
$P_{|g_{2},e,g_{2},0\rangle}(t_{f})+P_{|e,g_{2},g_{2},1\rangle}(t_{f})+P_{|g_{2},g_{2},e,0\rangle}(t_{f})+P_{|g_{2},g_{2},g_{2},1\rangle}(t_{f})$
where black areas correspond to approximately zero population. The
white dot shows specific values of $z_{0}$ and $d$ used in Fig.
\ref{pop1} to obtain a fractional STIRAP process.}\label{cont1}
\end{figure}
\begin{figure}
\includegraphics[width=7cm]{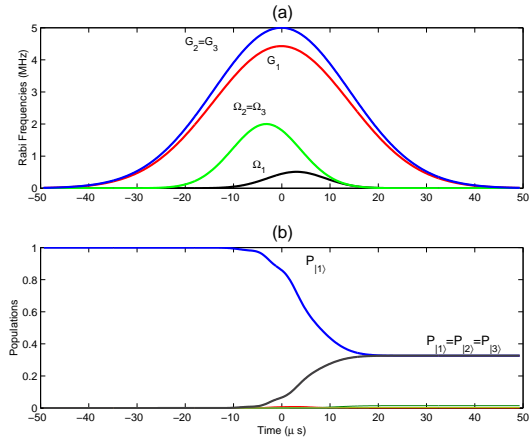}
\caption{(color online). (a): Rabi frequencies of the cavity-mode
and the laser field for three atoms in scheme 1, using the pulse
parameters of Fig. \ref{cont1}. (b): Time evolution of the
populations which represents a three-atom fractional-STIRAP. The
population of the states
$\{|e,g_{2},g_{2},0\rangle,|g_{2},e,g_{2},0\rangle\,|g_{2},g_{2},g_{2},1\rangle,|g_{2},g_{2},e,0\rangle\}$
is almost zero during the whole dynamics.} \label{pop1}
\end{figure}
The appropriate values of $z_{0}$ and $d$ that lead to the
f-STIRAP process can be extracted from a contour plot of the final
population $P_{|g_{1},g_{2},g_{2},0\rangle }(t_{f}):=|\langle
g_{1},g_{2},g_{2},0|\Phi (t_{f} )\rangle |^{2}$ as a function of
$z_{0}$ and $d$ that we calculated numerically (see Fig.
\ref{cont1}). The white dot in Fig. \ref{cont1}
 shows values of $z_{0}$ and $d$ to obtain an f-STIRAP process with the final populations of $P_{|g_{1},g_{2},g_{2},0\rangle
}(t_{f})\simeq P_{|g_{2},g_{1},g_{2},0\rangle}(t_{f})\simeq
P_{|g_{2},g_{2},g_{1},0\rangle}(t_{f})\simeq \frac{1}{3}$, and
zero population of the other states. Figure \ref{pop1}   shows (a)
the time dependent  Rabi frequencies of fractional STIRAP for
three atoms using the specific values of $z_{0}$ and $d$ in Fig.
\ref{cont1}, and (b) the time evolution of populations which shows
$\frac{1}{3}$ population for the states $
|g_{1},g_{2},g_{2},0\rangle$, $|g_{2},g_{1},g_{2},0\rangle$,
$|g_{2},g_{2},g_{1},0\rangle$ and zero population for the states
$|e,g_{2},g_{2},0\rangle$, $|g_{2},e,g_{2},0\rangle$
,$|g_{2},g_{2},e,0\rangle$, $|g_{2},g_{2},g_{2},1\rangle$  at the
end of the interaction. This case corresponds to the generation of
the maximally  entangled W state (\ref{Phi-fin})
 by adiabatic passage.
\subsection{Scheme 2}\label{scheme2}
In the second scheme, the system is taken to be initially in the
state $|\Phi ( t_{i} )\rangle=|g_{2},g_{2},g_{2},1\rangle$. Using
multilevel  STIRAP process \cite{unanyan}, and Rabi frequencies of
the fields as $G_{1}(t)=G_{2}(t)= G_{3}(t)= G(t)$,
$\Omega_{1}(t)=\Omega_{2}(t)=\Omega_{3}(t)=\Omega(t)$, We try to
transfer the final population of the system to the target W state.
In this case, the  dark state of $H^{\text{eff}}(t)$ is:
\begin{equation}\label{dark2}
|D(t)\rangle=C\Big(|g_{1},g_{2},g_{2},0\rangle
-\frac{\Omega}{G}|g_{2},g_{2},g_{2},1\rangle+|g_{2},g_{1},g_{2},0\rangle+|g_{2},g_{2},g_{1},0\rangle\Big),
\end{equation}%
 The possibility of decoherence-free generation of W state arises
from the following behavior of this dark  state:
\begin{subequations}
\begin{eqnarray}\label{b1}
  &&\lim_{t\rightarrow t_{i}} G(t)=0,\qquad |D(t_{i})\rangle= |g_{2},g_{2},g_{2},1\rangle,\\\label{b2}
  \nonumber\\&&\lim_{t\rightarrow t_{f}}\Omega(t)=0,\qquad
  |D(t_{f})\rangle=\frac{1}{\sqrt{3}}\big(
|g_{1},g_{2},g_{2},0\rangle+|g_{2},g_{1},g_{2},0\rangle+|g_{2},g_{2},g_{1},0\rangle\big).
\end{eqnarray}
\end{subequations}
\begin{figure}
\includegraphics[width=7cm]{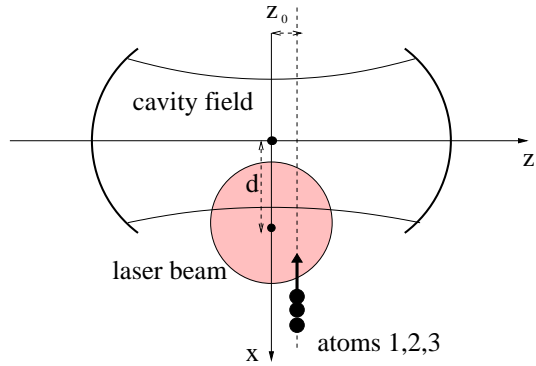}
\caption{The proposed geometry of the cavity and the laser fields
in $xz$ plane as well as the trajectory of the atoms for
generation of the atomic W state in scheme 2. The three atoms
initially in the ground state $|g_{2}\rangle$ arrive
simultaneously at the center of the cavity, initially in the state
$|1\rangle$. These atoms encounter the sequence laser-cavity on
the line $z=z_{0}$.} \label{xy2}
\end{figure}
\begin{figure}
\includegraphics[width=7cm]{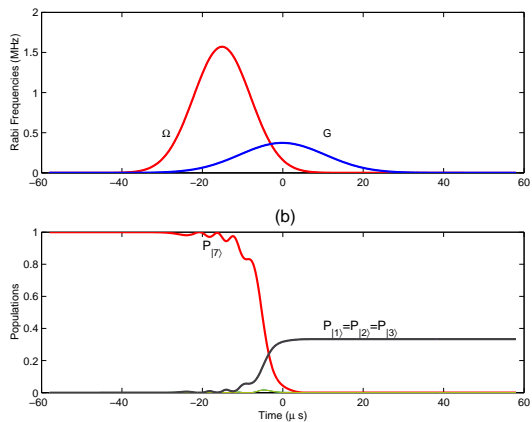}
\caption{(color online). (a): Rabi frequencies of the cavity-mode
and the laser fields (see Eqs. (\ref{G2}, \ref{G3})) for three
atoms with the pulse parameters as $v=2$ m/s,
$W_{L}=20\protect\mu$m, $W_{C}=30~\mu $m, $\lambda =780$ nm,
$\Omega _{0}=20$ MHz, $G_{0}=0.5$ MHz, $z_{0}=31.9~\mu$m,
$d=30~\mu$m. (b): Time evolution of the populations which
represents a three-atom STIRAP, and final population of $1/3$ in
the states
$\{|g_{1},g_{2},g_{2},0\rangle,|g_{2},g_{1},g_{2},0\rangle,
|g_{2},g_{2},g_{1},0\rangle\}$. We observe that the states
$|e,g_{2},g_{2},0\rangle,|g_{2},e,g_{2},0\rangle,|g_{2},g_{2},e,0\rangle,|g_{2},g_{2},g_{2},1\rangle
$ are not populated during the whole dynamics.} \label{pop2}
\end{figure}
In this scheme,  we consider a situation where the three atoms,
initially in the ground state $|g_{2}\rangle$, are going
simultaneously to interact with the  laser and cavity-mode fields
on the line $z=z_{0}$ (see Fig. \ref {xy2}), but through a
multilevel STIRAP process. The initial state of the cavity mode,
despite the first scheme, is $|1\rangle$.
 The atoms encounter time-dependent and delayed Rabi frequencies
given by
\begin{subequations}
\begin{eqnarray}
 \label{G2} G_{1}(t)&=&G_{2}(t)=G_{3}(t)=G(t)=G_{0}~e^{-(vt)
^{2}/W_{C}^{2}}\cos \left( \frac{2\pi z_{0}}{\lambda
}\right),\\\label{G3}
 \Omega_{1}(t)
&=&\Omega_{2}(t)=\Omega_{3}(t)=\Omega(t)=\Omega
_{0}~e^{-z_{0}^{2}/W_{L}^{2}}~e^{-(vt+d) ^{2}/W_{L}^{2}}.
\end{eqnarray}
\end{subequations}
 By multilevel STIRAP process \cite{unanyan}, with the sequence of laser-cavity (see Fig. \ref{pop2}),
 we can transfer the population from the initial state
$|g_{2},g_{2},g_{2},1\rangle$ to the final state (\ref{Phi-fin}).
Since the cavity-mode state factorizes and is left in the vacuum
state, there is no projection noise when one traces over the
unobserved cavity field, and the cavity is ready to prepare
another atomic entangled state.
\section{conclusion}
 we have proposed two robust and decoherence-free schemes to
generate atomic entangled W state, using the f-STIRAP and
multilevel STIRAP techniques in $\Lambda$-systems. These schemes
are robust with respect to variations of the velocity of the atoms
$v$, of the peak Rabi frequencies $G_{0},\Omega_{0}$ and  of the
field detunings. The first scheme is not robust with respect to
the parameters $d,z_{0}$, describing the relative positions of the
laser beam and the cavity, shown in Fig. \ref{xy1}, and the second
scheme requires initially one photon in the cavity mode. Assuming Gaussian pulse profiles for $\Omega (t)$ and $G(t)$ of widths $%
T_{L}=W_{L}/v$ and $T_{C}=W_{C}/v$ respectively, the sufficient
condition of global adiabaticity \cite{stirap} is $\Omega
_{0}T_{L},~~G_{0}T_{C}\gg 1$. Our schemes can be implemented in an
optical cavity with $G_{0}\sim\kappa\sim\Gamma$. In the second
scheme, $G_{0}$ can be smaller than $\Omega_{0}$
($G_{0}=\frac{1}{40}\Omega_{0}$ in Fig. \ref{pop2}), but in the
first one, the necessary condition to suppress the cavity
decoherence is $G_{0}\gg\Omega_{0}$ which is satisfied in practice
for $G_{0}\sim3\Omega_{0}$ (see Fig. \ref{pop1}). In scheme 1,
decoherence channels are suppressed during the whole evolution of
the system, while in the second scheme, decoherence is only
related to the initial photon of the cavity mode. In these
schemes, as opposed to other schemes,  we do not need to fix the
atoms inside the optical cavity. We can generalize this process to
generate $N$-atom W state with $N>3$.
\begin{acknowledgments}
M. A-T. wishes to acknowledge the financial support of the MSRT of
Iran and Urmia University. We thank H. R. Jauslin for helpful
discussions.
\end{acknowledgments}
\bibliography{paper7}
\end{document}